\documentclass[journal]{IEEEtran}
\usepackage[utf8]{inputenc}
\usepackage{amsmath,graphicx,amssymb,amsthm}
\usepackage{xcolor}

\begin{document}
\title{Towards Cooperative Federated Learning \\ over Heterogeneous Edge/Fog Networks}%Towards Cooperative and Generalized Federated Learning Over The Edge}
\author{Su Wang,~\IEEEmembership{Student~Member,~IEEE}, Seyyedali Hosseinalipour,~\IEEEmembership{Member,~IEEE}, Vaneet~Aggarwal,~\IEEEmembership{Senior~Member,~IEEE}, Christopher G. Brinton,~\IEEEmembership{Senior~Member,~IEEE}, David~J.~Love,~\IEEEmembership{Fellow,~IEEE}, Weifeng~Su,~\IEEEmembership{Fellow,~IEEE}, and Mung~Chiang,~\IEEEmembership{Fellow,~IEEE}
\thanks{S. Wang, V. Aggarwal, C. Brinton, D. Love, and M. Chiang are with Purdue University, IN, USA e-mail: \{wang2506, vaneet, cgb, djlove, chiang\}@purdue.edu.}
% \thanks{S. Wang, C. Brinton, D. Love, and M. Chiang are with the Elmore School of ECE at Purdue University, IN, USA e-mail: \{wang2506, cgb, djlove, chiang\}@purdue.edu.}
% \thanks{V. Aggarwal is with the School of Industrial Engineering at Purdue University, IN, USA e-mail: vaneet@purdue.edu}
\thanks{S. Hosseinalipour and W. Su are with the University at Buffalo--SUNY, NY, USA e-mail: \{alipour, weifeng\}@buffalo.edu.}} %the Department of Electrical Engineering at 
\maketitle

\begin{abstract}
    Federated learning (FL) has been promoted as a popular technique for training machine learning (ML) models over edge/fog networks. 
    Traditional implementations of FL have largely neglected the potential for inter-network cooperation, treating edge/fog devices and other infrastructure participating in ML as separate processing elements. Consequently, FL has been vulnerable to several dimensions of network heterogeneity, such as varying computation capabilities, communication resources, data qualities, and privacy demands.
    % , which can enhance and push edge/fog networks beyond a simple sum of their constituent devices' capabilities. 
    % By neglecting cooperative opportunities, current FL implementations separate edge/fog network devices and infrastructure from each other, and thus suffer from diverse forms of network heterogeneity, such as varying computation capabilities, communication resources, data qualities, and privacy demands. 
    {\color{black}We advocate for cooperative federated learning (CFL), a cooperative edge/fog ML paradigm built on device-to-device (D2D) and device-to-server (D2S) interactions. Through D2D and D2S cooperation, CFL counteracts network heterogeneity in edge/fog networks through enabling a model/data/resource pooling mechanism, which will yield substantial improvements in ML model training quality and network resource consumption.}
    % {\color{black} We advocate for cooperative federated learning (CFL), a cooperative edge/fog paradigm built on device-to-device (D2D) and device-to-server (D2S) interactions.} 
    % {\color{black} Through D2D and D2S cooperation, CFL counteracts network heterogeneity in edge/fog through enabling a model/data/resource pooling mechanism over the network, which yields substantial improvements in ML model training quality and network resource consumption.} 
    %savings in network resources through careful design of the associated cooperation mechanics.
    % Through careful design of cooperation mechanics among network devices and infrastructure, cooperative edge/fog can be further exploited to integrate unlabeled data and heterogeneous device privacy. 
    We propose a set of core methodologies that form the foundation of D2D and D2S cooperation and present preliminary experiments that demonstrate their benefits. We also discuss new FL functionalities enabled by this cooperative framework such as the integration of unlabeled data and heterogeneous device privacy into ML model training. %privacy restriction diversity 
    %showing the promise of a cooperative edge/fog paradigm for network intelligence applications. 
    % For each pillar, we propose a set of core technologies, embedded within a network schematic, to enable ML training  % present a network schematic embedded with a set of proposed technologies designed to enhance resource and ML training utility. 
    % We also demonstrate how such edge/fog cooperation can enable FL to adapt to scenarios involving unlabeled data. 
    Finally, we describe some open research directions at the intersection of cooperative edge/fog and FL. 
\end{abstract}

\section{Introduction} \label{sec:intro}
\noindent 
{\color{black} Recently, much attention has been given to the implementation of data analytics and machine learning (ML) techniques at the network edge to handle the complexity of emerging Internet of Things (IoT) services, ranging from user-oriented (e.g., object recognition) to network-oriented (e.g., signal classification) applications~\cite{salau2022recent}.} 
IoT devices are now capable of gathering data from various sources, connecting to the internet, and performing computation tasks. 
Collectively, they form edge/fog networks capable of producing machine intelligence insights. %can be developed by the collaboration of many such IoT devices. - only through cooperation
% that are greater than the sum of their parts, as 
%are created by the joining of these IoT devices together.  
% Together, these devices are greater than the sum of their parts form IoT networks that yield useful analytics and machine intelligence insights
% Formed by large quantities of such IoT devices, IoT networks are notable for having many individual edge device each contributing relatively little but together producing useful analytics and machine intelligence insights. 
{\color{black}Traditionally, data insights were produced via centralized computing, where network devices send all of their local measurements to a single central server for ML training. Such methods led to system-wide latency and resource inefficiencies as a result of data transmissions from edge devices to the server, for centralized ML tasks specifically. These limitations have led to the emergence of distributed ML techniques and in particular federated learning (FL).}

\begin{figure*}[t!]
\includegraphics[width=.98\textwidth]{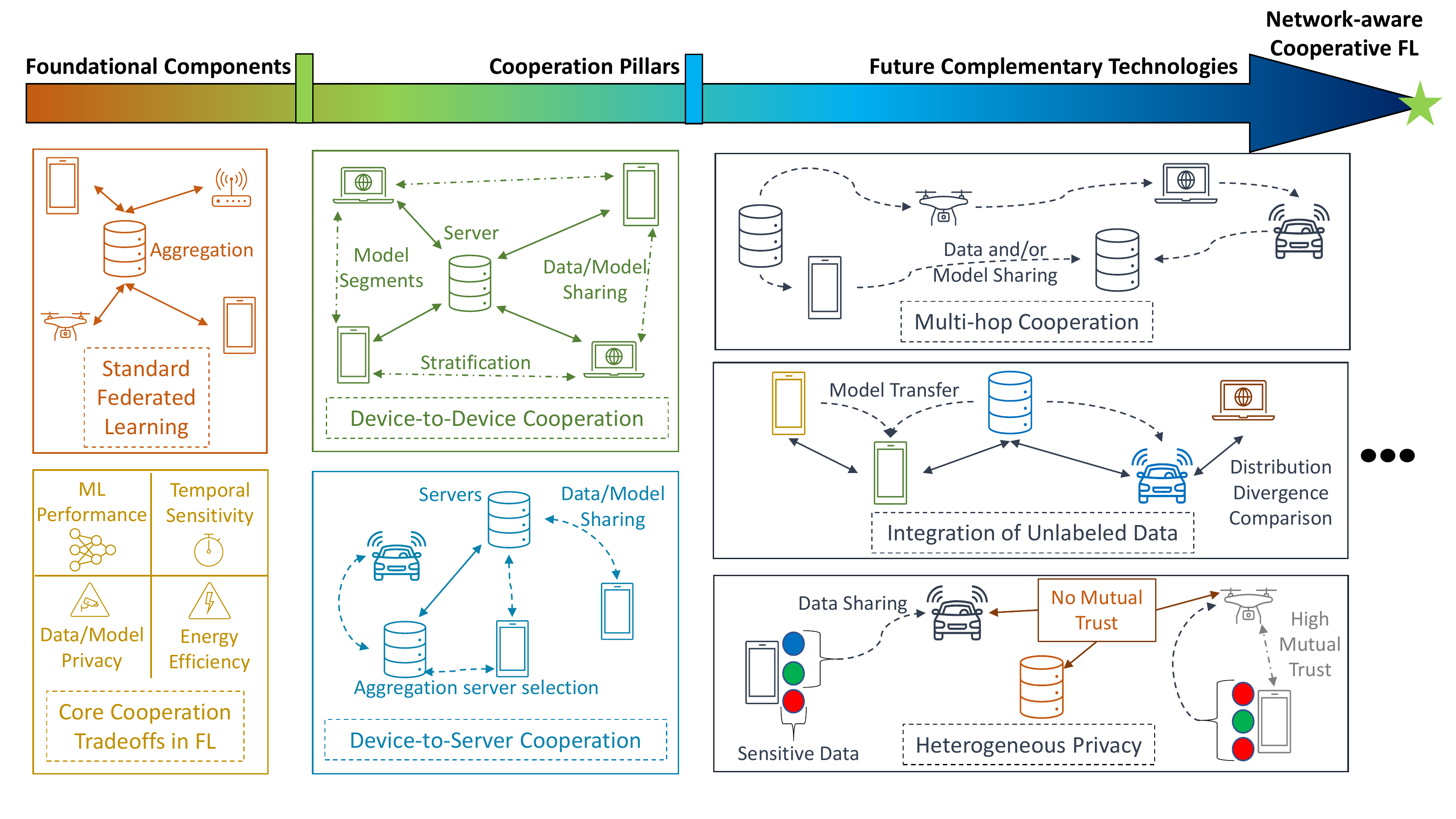}
\centering
\caption{{\color{black} The progression from standard FL towards network-aware cooperative FL. Starting from standard FL and its associated trade-offs, we develop the pillars of cooperative FL frameworks, namely device-to-device and device-to-server cooperation. 
We envision that future work towards integrating collaboration into FL involves ideas such as multi-hop cooperation, integration of unlabeled data, and heterogeneous privacy.}}
\label{fig:main_diagram}
\vspace{-3mm}
\end{figure*}

{\color{black}Standard FL shifts the processing portion of ML training from the server to the edge/fog devices~\cite{kairouz2021advances}. As shown in the upper left corner of Fig.~\ref{fig:main_diagram}, it involves a ``star" server-to-device communication topology, inside of a three-part cyclical process: (i) edge devices independently and locally train an ML model, (ii) ML models are sent to the central server for global aggregation, and (iii) the server synchronizes devices' ML models into an aggregated ML model called the global model.}
% The global aggregation mechanism of standard FL is the only form of network cooperation among network elements. 
While standard FL features global aggregations, this is the only form of cooperation among network elements. 
Inter-device and inter-network communications - key features of IoT networks - can also facilitate cooperation and are  missed opportunities in FL. For example, direct device-to-device (D2D) communication links that are otherwise underutilized could be employed for faster and communication-efficient ML model training~\cite{su2021secure}. 
%the underutilized network links in FL can be leveraged for 
% This leads to underutilized network links such as for device-to-device (D2D) cooperation - a missed opportunity that is especially pronounced in IoT networks built on the power of inter-device and inter-network collaborations.  
% In standard federated learning, this process is the only form of edge device and/or network cooperation, and results in underutilized network resources and links. 
Traditional FL therefore does not exploit the full potential of cooperation in large-scale edge/fog networks. %on large-scale edge/fog networks are therefore not exploiting the full potential of their network. 

% {\color{black}We propose cooperative federated learning (CFL), a cooperative edge/fog paradigm that jointly orchestrates the resources at devices, servers, and network infrastructure to enhance FL and its core trade-offs as shown in Fig.~\ref{fig:main_diagram}.}
{\color{black}We propose cooperative federated learning (CFL), a cooperative edge/fog ML paradigm that jointly orchestrates device, server, and network infrastructure resources to enhance FL while considering its core trade-offs, as shown in Fig.~\ref{fig:main_diagram}.}
{\color{black}CFL extends the notion of cooperation to address the key missed opportunities in standard FL, which are summarized below:}
% in standard FL to address a set of key missed opportunities summarized below:} 
\begin{enumerate}
    \item Edge devices with powerful local processors or small local datasets idly wait for network stragglers to finish training~\cite{reisizadeh2022straggler}.  %complete their ML training tasks
    \item Powerful network infrastructure elements such as edge/fog servers are underutilized in FL~\cite{mai2022federated}. %are left idle in the FL process
    \item Edge devices without direct connectivity to the central server are neglected during ML model training and synchronization processes~\cite{kairouz2021advances}. %\cite{li2020federated}. 
    \item IoT devices, which may have diverse privacy requirements~\cite{lin2021friend}, are all discouraged from sharing data/models over the network. %and ML tasks
\end{enumerate}
These shortcomings are the result of prohibiting powerful devices, idle edge servers, and network infrastructure from helping computationally weak and/or overburdened devices in FL. 
% At the same time, edge/fog environments are notable for containing the above statistical and structural heterogeneity at a large-scale~\cite{li2022data}. 
{\color{black}Simultaneously, D2D and device-to-server (D2S) cooperation over such edge/fog networks has been shown to be feasible and beneficial to learning processes~\cite{kar2023offloading}.}
Well-designed cooperation mechanisms can thus unlock the full potential of edge/fog networks for FL, leading to (i) improved ML performance, (ii) energy efficiency, (iii) temporal efficiency (e.g., faster ML training), and (iv) diverse data/model privacy. %a broader range of data/model privacy management. 
%.e., less idle time during model aggregations, faster data processing, etc.)

\section{Cooperative Federated Learning}
% \noindent To maximally exploit network resources to improve FL, edge/fog networks should leverage underutilized network elements (i.e., devices/servers/infrastructure) to help overloaded ones. %shoulder the burden of 
\noindent 
{\color{black} We propose cooperative federated learning (CFL), a novel paradigm that expands the dimensions of cooperation in FL beyond global aggregations. CFL develops inter-element cooperation mechanisms including selective data sharing, computation resource sharing, ML model sharing, and data distribution comparisons. 
FL driven by such multi-faceted cooperation better exploits the availability of links among network devices, servers, and infrastructure in contemporary edge/fog systems. Through such links, CFL unlocks the potential of cooperative edge/fog networks for ML.} 

% {\color{black} Unlike the state-of-the-art in FL~\cite{kairouz2021advances}, which has thus far only leveraged data offloading to improve the model training of FL, our vision for CFL involves a broader look at cooperation in FL, including D2D- and D2S-driven data, model, and resource cooperation.} 
{\color{black}Whereas the state-of-the-art in FL~\cite{kairouz2021advances} has mostly considered data offloading as a mechanism for improving local statistical properties in FL, our vision for CFL involves a broader look at cooperation in FL, including D2D- and D2S-driven data, model, and resource cooperation.}
{\color{black}
These proposed cooperation technologies aim to improve the balance among the trade-offs in FL shown in Fig.~\ref{fig:main_diagram}. For example, intelligent cooperation can lead to better ML model performance with less energy usage and system delay.}
% These proposed cooperation technologies can change the balance of the trade-offs of FL, shown in Fig.~\ref{fig:main_diagram}. For example, cooperation can lead to better ML model performance with less energy usage and system delay.}

% We term such inter-element assistance as cooperation, which encompasses data sharing, computation resource sharing, localized ML model transfers, data distribution comparisons, and aggregation scaling. 
% FL driven by such cooperation better exploits the links among network devices, servers, and infrastructure and through such links unlocks the potential of cooperative edge/fog networks. We call this paradigm cooperative federated learning (CFL). 

Specifically CFL (i) leverages D2D links for data and model sharing, which we term D2D cooperation, and (ii) leverages D2S links to incorporate edge servers, routers, and other network infrastructure into the FL ecosystem through data processing and ML model transmission tasks. 
{\color{black} We consider data transfers in CFL noting that while some applications of FL (e.g., healthcare analytics) discourage data sharing, other applications have milder data privacy restrictions, especially when the data is generated with ML as the primary purpose (e.g., FL for self-driving vehicles with sensor measurements).}

Combined, D2D and D2S cooperation form the two pillars of CFL, which together enable many complementary technologies,  we only examine a few of which for brevity. As depicted in Fig.~\ref{fig:main_diagram}, we examine (i) multi-hop cooperation due to its key influence on improved resource efficiency (i.e., energy efficiency and temporal sensitivity), (ii) integration of unlabeled data as it enhances ML performance for devices, and (iii) devices heterogeneous privacy demands because it focuses on the data/model privacy aspects of CFL.  % both devices with and without labeled data

\subsection{Overarching Technologies for CFL} \label{ssec:design_intro}
% CFL, through its two pillars of D2D and D2S cooperation, addresses the key missed opportunities of standard FL. %the two key pillars of CFL
In the following, we explain how D2D and D2S cooperation exploit the network characteristics inherent in edge/fog networks, and fulfill the missed opportunities of FL. % and establish the foundation for emerging applications in fog networking. %further and future development towards network-aware CFL. 
%Within each section, we explain how these key pillars of CFL fulfill the missed opportunities of FL and how they establish the foundation for further and future development towards network-aware CFL. 

\subsubsection{Device-to-device (D2D) cooperation}
In edge/fog networks, devices are heterogeneous statistically (i.e., different dataset characteristics) and structurally (i.e., varying computation/communication capabilities). %cpu processing capability and local energy availability). %underlying data distributions and local data quantity 
In standard FL, such heterogeneity leads to isolated resource-abundant and resource-scarce devices, some of which may introduce straggler effects and delay ML model aggregations. In the worst-case, resource-scarce devices may be unable to complete ML model training iterations, possibly due to insufficient battery, or result in the existence of unused local data, as model training in straggler devices use smaller batches of data. % are used for model training in straggler devices. 
To cope with device heterogeneity, we exploit D2D cooperation as discussed below.

% Rather than treating network heterogeneity as a pure disadvantage, we can unlock the potential arising from edge/fog network heterogeneity via D2D cooperation. %, as explained below. 
% In the following, we detail how our proposed technologies enable D2D cooperation to unlock the benefits arising from edge/fog network heterogeneity in the following sections. 

\textbf{Cooperation as resource pooling.} Cooperative edge/fog networks can reallocate the intensity of local ML training by leveraging D2D links for data transfers from resource-scarce to resource-abundant devices. Through this process, the impact of stragglers (i.e., resource-scarce devices) on ML model training is then reduced.
% %, in effect yielding a resource pooling of data and processing power. 
% Data transfers enable resource-abundant devices to shoulder part of the ML training burden from resource-scarce devices, 
% reducing the straggler effects caused by resource-scarce devices. 
% As a result of data transfers, resource-abundant edge devices are able to shoulder part of the ML training burden from resource-scarce devices, reducing the straggler effects caused by resource-scarce devices. 
Simultaneously, D2D cooperation shifts the burden of ML model training to devices with energy-efficient processors, and thus can lead to network energy savings. %more of the

\textbf{D2D driven model offloading.}
Similar to data and subsequent ML training offloading, D2D cooperation can mitigate some of the model aggregation overhead. Rather than only transmitting local ML models to the aggregation server, devices can transfer different parts/chunks of their model to their neighboring devices. In this way, those devices that are far away from the server or have limited communication resources (e.g., limited bandwidth) can communicate with their neighboring devices.
% layers of their local model parameters to their neighbor devices. 
Devices that receive models from their neighbors then combine received models with their local one and then transmit these partially combined ML parameters to the server. 
% subsequently send the combined parameters to the aggregation server. 
Thus, D2D model offloading can reduce the number of devices engaging in resource intensive uplink transmissions.

% Due to this diversity in network composition, edge and fog devices will vary with respect to local data storage availability, computational power in terms of cpu processing power and cpu cycle effectiveness, communication cost modeling as channel conditions will vary for device pairs, energy availability, and underlying data distributions. 
% Proper design of inter-device collaboration enables many network energy saving opportunities, e.g., controlling cpu clock speeds and D2D data offloading so that efficient devices will process more data,  enabling multi-hop model parameter transmissions to take advantage of high quality and low cost communication channels, and finer grain control of the total duration of ML training. 
% D2D collaboration also influences fundamental design choices of standard FL, such as global aggregation scheduling and device sampling for aggregations. 
%To further integrate D2D collaboration and FL, this network device diversity will influence fundamental design choices of standard FL, 

\subsubsection{Device-to-server (D2S) cooperation}
Current FL research presumes conducting ML model training solely on the devices and neglects the underutilized network infrastructure elements such as edge servers. 
While edge servers may not gather data themselves, they can add value to the FL process owing to their  powerful local processors and dedicated communications equipment which can be exploited through network collaboration. % to enhance the FL process. 
% Especially so given the more powerful processors and dedicated communications equipment at edge servers and routers. 
%leaves the potential of network infrastructure elements such as edge servers and routers untapped. 
We will refer to all cooperation, aside from global model aggregations, between devices and edge servers as D2S cooperation. 
Two potential use cases of edge servers are provided below. 
% We next highlight two potential use cases of edge servers that can improve the FL process. 
%(i) computational resources, (ii) model aggregation gateways, and (iii) flexible data caches. 

\textbf{Computational resources.}
D2S cooperation allows resource-scarce devices to transfer their local data to a physically stable and computationally powerful edge server. These edge servers then function similarly as a resource-abundant device, enabling the network to process more training data and lessening  straggler effects. 

% \textbf{Model aggregation gateways.}
% As edge servers often contain dedicated and powerful communications equipment, they can reroute edge devices' transmissions at global aggregations. As a result, edge devices can transmit their local ML models to their nearby edge servers, which can subsequently transmit partially combined ML models from nearby devices to the aggregation server. Such D2S cooperation saves energy for devices as it bypasses higher powered long distance communications from devices to the aggregation server. 

% they are well placed to receive and subsequently transfer the ML model parameters from many nearby edge devices to the central or main aggregation server. 
 
% So model parameter transfer on D2S links reduces the total quantity of devices communicating to the central server, thereby decreasing overall network communication interference and mitigating channel congestion. 
% Edge servers can thus act as gateways to the central or main aggregation server. 

\textbf{Flexible data caches.}
Using the data they receive from nearby devices, edge servers can act as local data caches, which can carry globally representative distributions of high quality data, to mitigate the impact of non-i.i.d. data across the network. 
Additionally, this functionality enables better tracking of the distribution shifts in the data via comparing old data with newly arriving data at the edge servers, which enables more informative decisions on ML model training. %the knowledge of which can be used to take more informative decisions on ML model training.
% assessments of the time-varying effectiveness of ML models, which is especially useful in real-world dynamic networks where devices' underlying data distributions change over time.  %through changes in underlying data distribution

\subsection{Enhancing the Core Properties of CFL} %complementary technologies
% The above proposed technologies for D2D and D2S cooperation form a basis from which sophisticated applications and technologies can be developed. 
While many applications/extensions are possible from the foundation of D2D/D2S cooperation, we focus on a subset of techniques that can enhance the core trade-offs in CFL. We present high-level explanations of (i) multi-hop cooperation due to its benefits for resource efficiency (energy efficiency and time sensitivity), (ii) integration of unlabeled data as it extends FL to benefit a wider range of devices (e.g., devices with unlabeled data), and (iii) heterogeneous privacy for its enhancements to data/model privacy. 

% we focus on high-level explanations of (i) multi-hop cooperation, (ii) integration of unlabeled data, and (iii) heterogeneous privacy. 

\textbf{Multi-hop D2D and D2S cooperation.}
{\color{black}Multi-hop D2D and D2S cooperation refers to extending the above concepts developed for single-hop D2D and D2S cooperation to multiple, sequential links in between devices.} 
% can be extended from the above concepts for single-hop D2D and D2S cooperation. 
This envisioned technology can greatly improve the resource efficiency (i.e., less energy consumption and/or faster model training) of FL. In particular, multi-hop cooperation enables greater connectivity/reach from resource-scarce to resource-abundant edge devices or servers. 

% As a result, data/model offloading can be done more efficiently than in single-hop scenarios, leading to greater resource efficiency . 
% enable more devices that are normally incapable of contributing to FL process (due to to physical distance from the aggregation server or complete lack of local processing power) to transmit their local model parameters or data to nearby edge devices/servers to overcome these challenges.

\textbf{Integration of unlabeled datasets.} D2D and D2S cooperation can better represent and facilitate the contribution of devices with  partially labeled or unlabeled datasets in FL. % to be better represented or even contribute to the ML training process. 
{\color{black}Unlabeled data refers to data samples that have not been tagged with a ground-truth, e.g., images taken by cameras mounted on smart cars without pre-assigned or pre-identified types of objects within the image.} 
In standard FL, only devices with labeled data are engaged in ML model training. Consequently, edge devices with unlabeled datasets are unlikely to have their data properly represented at the global ML model and so are likely to suffer from poor ML performance. 
% We explain the mechanics in-depth in our discussions of future work, but, at a high level, 
Roughly speaking, D2D and D2S cooperation can enable approximations of the local data distribution of each device at its neighboring devices/network elements, even if the device has fully or mostly unlabeled data.
% , through cooperative data transfers across the network. 
Subsequently, the type of distributed learning method being applied can be tuned to improve ML performance across the network. 

\textbf{Heterogeneous privacy.} 
{\color{black}Similar to statistical (i.e., data-level) and structural (i.e., computation/communication resources) differences, edge/fog devices also exhibit heterogeneity with respect to their privacy needs.
Heterogeneous privacy needs will motivate selective D2D and D2S cooperation.
For example, D2D cooperation can involve sharing sensitive data only among mutually trusted devices (e.g., edge devices belonging to the same user or family), while among untrusted neighbors, this sharing can be limited to sharing insensitive data or even prohibited completely.
This is one of the future complementary technologies depicted in Fig.~\ref{fig:main_diagram}. With such methods in place, D2D/D2S cooperative technologies can improve the resource efficiency and ML performance while meeting data/model privacy requirements of edge/fog network elements.}

\subsection{Towards Network-Aware CFL}
% Originating from the pillars of D2D and D2S cooperation, we can develop complementary technologies, which all follow the paradigm of network-aware CFL. %This proposed paradigm relies on cooperative edge/fog networks to enhance standard FL 
As depicted in Fig.~\ref{fig:main_diagram}, our vision for CFL relies on cooperative edge/fog networks to enhance standard FL on (i) ML performance - the effectiveness of the ML model trained by the network, (ii) energy efficiency - the network-wide accumulated energy expenditure on data processing, and data/model communication, (iii) temporal efficiency - the total time including idle time consumed by the FL process, and (iv) data and model privacy - the heterogeneous privacy needs in large-scale edge/fog networks. 
% The design considerations outlined above motivate network-aware CFL, a novel paradigm that relies on cooperative edge/fog networks to enhance standard FL on four fundamental fronts: (i) ML performance - the effectiveness of the ML model trained by the network, (ii) energy efficiency - the network-wide accumulated energy expenditure on data processing, and model and data communication, (iii) temporal efficiency - the total time including idle time consumed by the FL process, and (iv) data and model privacy - the heterogeneous privacy needs in large-scale edge/fog networks. 
Methodologies that develop network-aware CFL must carefully balance their contributions to these four coupled elements, which contain design trade-offs. % which are coupled together and thus contain design trade-offs. 

% So, we first establish an analytical foundation depicted in Fig.~\ref{fig:main_diagram} that characterizes network-aware cooperative FL by four fundamental components: (i) ML performance - the classification effectiveness of the ML model trained by the network, (ii) energy efficiency - the network-wide accumulated energy expenditure on data processing, model and data communication, and idling, (iii) temporal sensitivity - measuring time consumption of the ML training and associated processes as a network resource, and (iv) data and model privacy - user controlled preservation of data and model integrity. 
% While these four aspects have some overlapping effects, for instance the most energy efficient method may not generate the best ML model performance, methods that move towards developing network-aware cooperative FL must consider these four elements explicitly. 

% After standard FL and the core trade-offs of cooperation, 
The first layer of network-aware CFL consists of the pillars of D2D and D2S cooperation and their core mechanisms. %The first layer of development consists of the fundamental mechanisms of device-to-device (D2D) and device-to-server (D2S) cooperation relative to the four guiding trade-offs/principles in a standard FL network. %must consider 
D2D and D2S cooperation can leverage data and model transfers to cope with device heterogeneity. % (e.g., varying computational power). 
{\color{black}For example, data offloading through D2D links can yield energy savings, and ML model routing through D2S links can mitigate straggler effects.}

With our established frameworks of D2D and D2S cooperation, we can then develop complementary technologies to enhance CFL along the four design considerations of (i) ML performance, (ii) energy efficiency, (iii) temporal sensitivity, and (iv) data/model privacy.
% Subsequently, we can extend the benefit brought by D2D and D2S cooperation along the four core CFL properties of (i) ML performance, (ii) energy efficiency, (iii) temporal sensitivity, and (iv) data/model privacy through proposed complementary technologies, 
In Fig.~\ref{fig:main_diagram}, we depict three sample complementary techniques, with each technique primarily improving one aspect of CFL. For instance, multi-hop collaborations such as those seen in industrial IoT~\cite{li2022data} enhance resource (energy and delay) efficiency, integration of unlabeled data such as those in autonomous driving~\cite{luo2021self} can improve ML performance, and heterogeneous privacy as seen in social trust~\cite{lin2021friend} offers an alternative approach to data/model privacy. 
% such as multi-hop collaborations in large-scale IoT~\cite{li2022data}, unlabeled datasets in autonomous driving~\cite{luo2021self}, or heterogeneous data privacy by connecting FL with social trust aspects~\cite{lin2021friend} as depicted in the complementary technologies layer in Fig.~\ref{fig:main_diagram}. 
We next develop D2D and D2S cooperation as the two pillars of network-aware CFL, and, as future work, explain how complementary technologies can enhance them. 

\section{Network-aware D2D Cooperation} %Multiple Aspects of D2D Collaboration} %developing network-aware D2D collaboration
\label{sec:psl_extensions}
\begin{figure}[t]
\includegraphics[width=.48\textwidth]{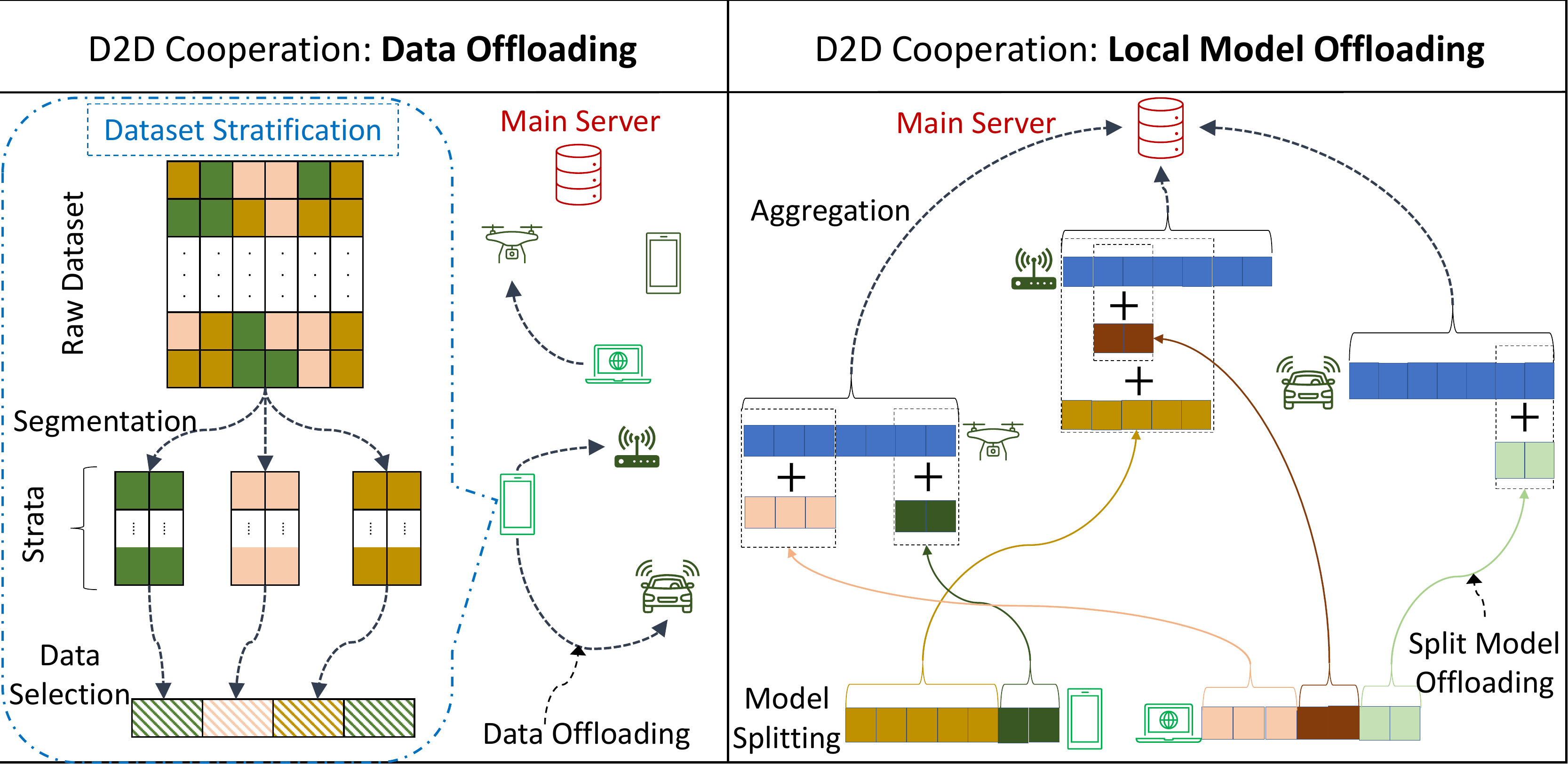}
\centering
\caption{Data and local model offloading form the basis of effective D2D cooperation. % in federated settings. %As part of effective D2D cooperation, 
We envision smart data offloading by edge devices through dataset stratification.
Additionally, we propose local model offloading, where devices offload segments of their local ML models to other devices. %nearby edge devices.
}
% , a technique in which edge devices can split and subsequently offload segments of their local ML models to nearby edge devices to save on communication resources.} 
% Segmenting local ML models and subsequently transmitting them to a range of edge devices transmission also offers an opportunity to increases the resolution of efficient D2D cooperation in the form of local model offloading.} 
\label{fig:PSLbig}
\end{figure}

\noindent
The first step towards network-aware CFL is to develop and maximize the benefits arising from D2D cooperation. 
{\color{black}Well-designed D2D cooperation can enable efficient orchestration of limited network resources, leading to improvements in the trade-offs of FL from Fig.~\ref{fig:main_diagram}.} % For example, efficient D2D data offloading  can improve energy consumption during ML training and reduce training delay by transferring data from resource-scarce devices to resource-abundant ones.} %to improve ML model training efficiency. %instance through data offloading from resource-scarce devices to resource-abundant ones in order to re-balance ML model training. 
To this end, we first introduce a set of core, overarching technologies to enable effective D2D cooperation, and thereafter propose future work on complementary technologies to further enhance D2D cooperation.

\begin{figure}[t]
\centering
\includegraphics[width=.48\textwidth]{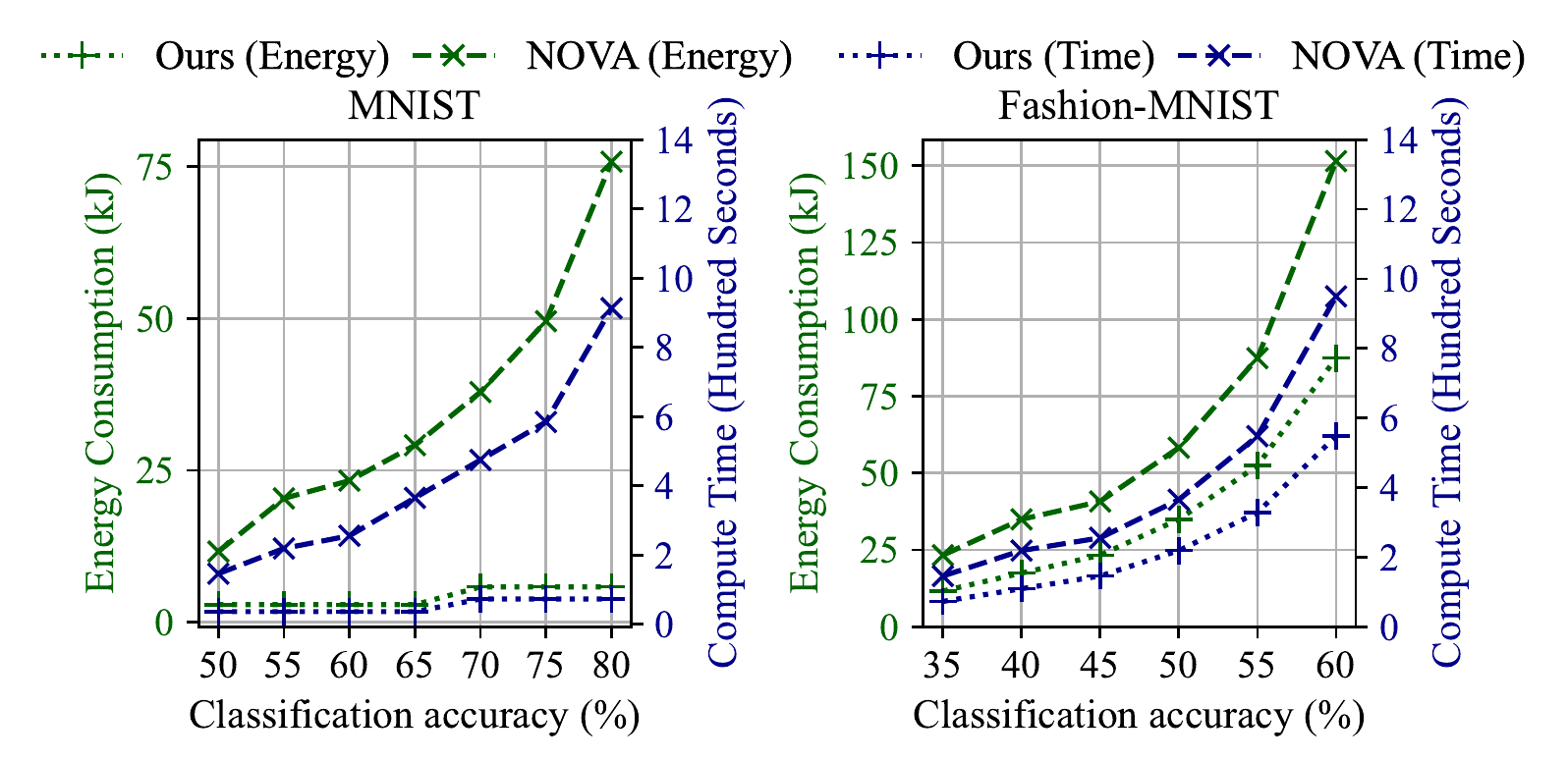}
\caption{{\color{black}Our D2D cooperation-driven method incurs both less compute time and energy consumption relative to the state-of-the-art $\mathsf{NOVA}$ methodology on two commonly used machine learning datasets (MNIST for numbers and Fashion-MNIST for clothes).}}
\label{fig:D2D_comparison}
\end{figure}

\subsection{Core Techniques for Effective D2D Cooperation} %Key Technologies Towards Effective D2D Collaboration}
Effective D2D cooperation improves network resource efficiency and ML model training through innovations in data and model offloading, which we propose in Fig.~\ref{fig:PSLbig}. % from resource-scarce devices to resource-abundant ones. 
At the data level, we propose dataset stratification, a method which clusters local datasets for higher quality data offloading. %segments vs clusters 
% technologies that jointly improve the resource consumption (energy and delay) and ML model performance. 
At the ML model level, we propose local model offloading, a technique that involves partial local ML model offloading to streamline efficient ML model aggregations. % local model segmentation and subsequent offloading to reduce resource consumption during ML model aggregations. 
% to sequentially orchestrate model aggregations through D2D model offloading. 
% We explain these proposed technologies in detail below, and provide a high-level overview in Fig.~\ref{fig:PSLbig}.

\subsubsection{Dataset Stratification}
After offloading data, the sending device (sender) continues local ML model training on a smaller local dataset, as keeping a copy of the transferred data leads to bias at ML model aggregations from counting the same data multiple times. %twice. %smaller vs lower-quality
% When offloading data, the sender should not use the offloaded subset of data for local ML training as otherwise it would lead to double counting at model aggregations and thus bias the global ML model. 
On the other end, the receiving device (receiver) may receive data that is unrepresentative of the data gathered at the sender. 
As a result, random/naive data offloading may hinder rather than help the ML model training, motivating dataset stratification. 

As depicted on the left subplot of Fig.~\ref{fig:PSLbig}, dataset stratification clusters datasets into strata (i.e., categories) based on task-dependent criteria. 
{\color{black}
Using the example of clothing recognition (i.e., the Fashion-MNIST dataset~\cite{hosseinalipour2022parallel}) for data collected by smartphones' cameras, each stratum may contain data belonging to a unique type of clothing (e.g., T-shirts or coats).
% The datasets of devices will contain datapoints from different broadcast sources. Through stratification, each strata can contain only data belonging to a unique type of broadcaster. 
Then, through sequential selection of the most representative data samples (i.e., those that are closest to the strata average) from the most populous strata, devices can offload datapoints which well-capture the distribution of their local dataset.}
% in signal source classification, each strata would contain only data belonging to a unique signal source designation, such as cell phone, sensor, or information server. 
% Then, through sequential selection of the most representative data samples (i.e., those that are closest to the strata average) from the most populous strata, devices can offload datapoints which well-capture the distribution of their local dataset.} 
In doing so, dataset stratification (i) keeps the distribution of the dataset at senders relatively intact, and (ii) enables receivers to receive a representative sample of data from senders. 
% offloads data by 
% enables (i) senders to selectively offload data so that their local ML training remains representative of their originally gathered data, and (ii) receivers to augment their local datasets with data representative of their neighbors' datasets. %(e.g., to reduce dataset heterogeneity across edge devices)

As an example, consider a smart car communicating with a drone. From its operation, the car has images of mostly stop signs and traffic lights, which the drone may not. Dataset stratification enables the car to transmit a small set of representative stop sign and traffic light pictures, without significantly distorting its local dataset distribution, to the drone. 
% smart cars gather data from different underlying environments. While one vehicle may have images of buildings and highways, another may have images of trees and small animals. These two vehicles may also have some overlapping data such as images of stop signs and traffic indicators. If these two vehicle were to meet, dataset stratification would enable the image sharing of buildings and highways for trees and small animals, i.e., the transmission of data useful to both vehicles, rather than overlapping images of stop signs. 

\subsubsection{Local Model Offloading}
% On the model cooperation side, we propose local model offloading,
We propose segmenting and sequentially transmitting the devices' local ML models at global ML model aggregations. {\color{black}Each segment of an ML model is a subset of model parameters. For example, in the case of neural networks, each segment may contain the parameters associated with a layer of the neural network. Local model offloading, depicted in Fig.~\ref{fig:PSLbig}, enables devices with limited access to the server (e.g., due to unsatisfactory channel conditions) to offload different segments of their local ML models to intermediary devices with a better accessibility to the server.}
% Local model offloading, depicted in Fig.~\ref{fig:PSLbig}, enables devices to offload different layers to different intermediary devices on the path to the central server.} % splitting devices' local ML models into varying segments, corresponding to different layers of a neural network, and subsequently transferring these different segments to intermediary devices on the path to the central server. }
These intermediary devices will combine their local ML models with their received partial ML models, leading to a set of partially aggregated parameters, which are then sent to the central server. 
In this way, the central server is able to perform the global aggregation using all ML model parameters while saving communication resources. 

% \subsubsection{ML Performance, Energy Consumption, and Delay Optimization}
% By combining these technologies into CFL, we are able to develop a joint ML performance, energy consumption, and delay optimization formulation. 
{\color{black}We have taken initial steps toward formalizing this D2D cooperation methodology in our prior work~\cite{hosseinalipour2022parallel}, including optimization formulation, theoretical results, and convergence proofs. 
To evaluate its potential benefits, we compare it against the algorithm Nova~\cite{wang2021novel} on two common datasets used to evaluate FL methods: MNIST (numbers) and Fashion-MNIST (clothes). Simulation results are shown in Fig.~\ref{fig:D2D_comparison}, where both methods train a two layer convolutional neural network (CNN) across a network of 10 devices. A detailed description of the computational infrastructure, wireless channel models, and models of energy consumption (including energy from both the communication and computation processes) can be found in~\cite{hosseinalipour2022parallel}.}
{\color{black}We compare against Nova~\cite{wang2021novel}, a recent and well-established method for aggregations involving heterogeneous training epochs across edge/fog devices.}
% To evaluate the impact of the above proposed technologies, we compare them against Nova~\cite{wang2021novel} on two common datasets used to evaluate FL methods: MNIST (numbers) and Fashion-MNIST (clothes), in Fig.~\ref{fig:D2D_comparison}. 
% In our simulation, both methods train a two layer CNN across a network of 10 devices. For this result, we used the same computational infrastructure, channel models, and energy modeling as in our prior work~\cite{hosseinalipour2022parallel}.} 
% {\color{black}We compare against Nova~\cite{wang2021novel} as it is the state-of-the-art method for aggregations involving heterogeneous training epochs across edge/fog devices.}
Our proposed CFL technology is seen to yield (i) consistent energy savings, and (ii) faster ML training times. 

\begin{figure}[t]
\includegraphics[width=.48\textwidth]{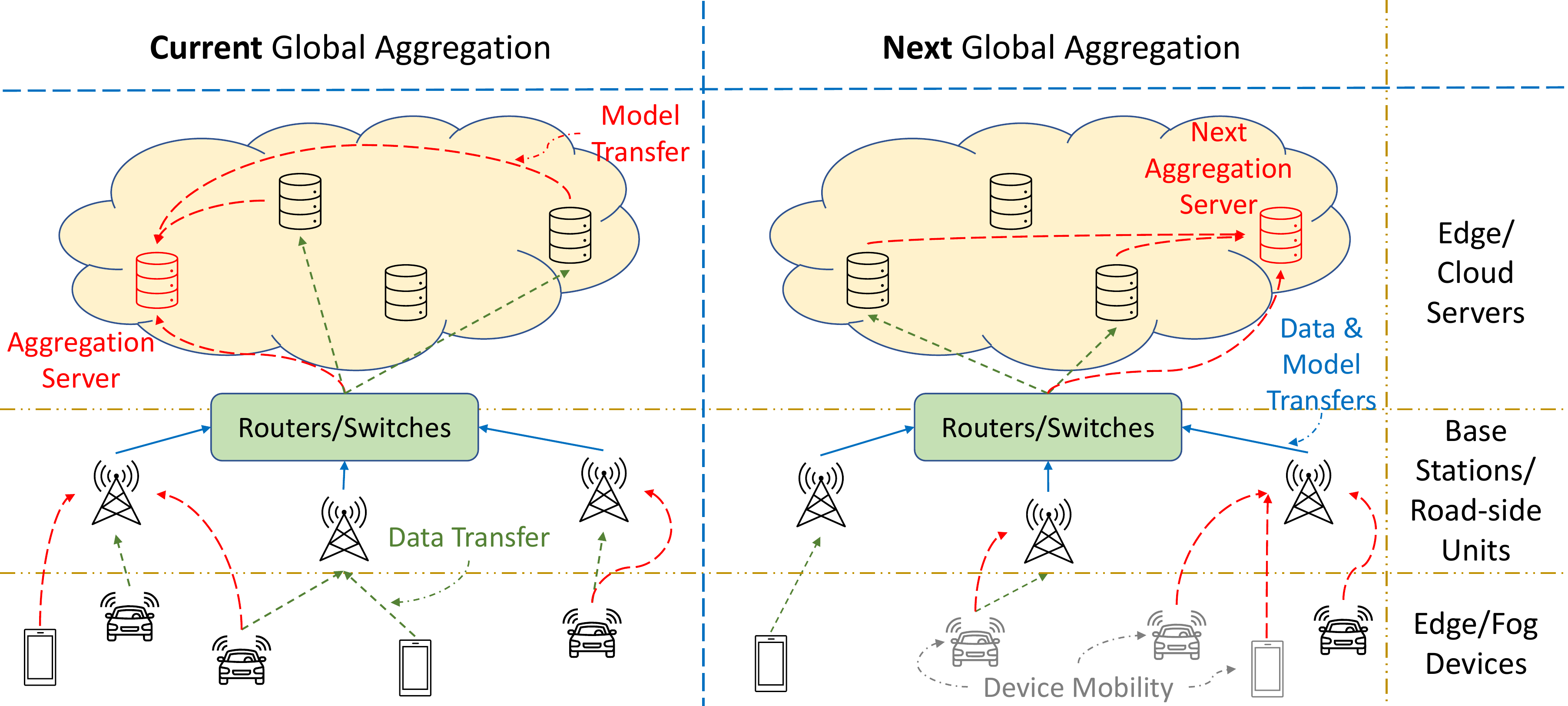}
\centering
\caption{Device-to-Server (D2S) cooperation via data/model transfers can improve the resource efficiency and performance of FL. %and the ML performance. 
Additionally, intelligent selection of the aggregation server can further reduce aggregation delay.} %, we can further reduce the aggregation delay.}%, especially edge/fog networks with device mobility. }
% To drive these benefits further, we can adjust flexibly adjust our selection of aggregation servers.} 
%{\color{black}D2S diagram. Plan is to better emphasize D2S collaboration and the time-varying central server for aggregation.} }
\label{fig:D2S}
\vspace{-3mm}
\end{figure}

\subsection{Future Development of Complementary Technologies}
Further extension of D2D cooperation can further enhance the trade-offs in FL and subsequently CFL depicted in Fig.~\ref{fig:main_diagram}. %developing its robustness to network changes or integration of multi-hop offloading topologies, which can enhance the core properties of CFL from Fig.~\ref{fig:main_diagram}. %ML performance, energy efficiency, %These complementary technologies 
{\color{black} A few open research directions are summarized below:} %We summarize the more interesting open research directions below: 
\subsubsection{D2D cooperation in non-stationary networks}
In non-stationary networks, devices will enter and exit the network, leading to varying network size, changing compute resource availability, and time-varying D2D links. 
Here effective cooperation should consider the physical stability of edge/fog devices to determine effective time-varying anchor devices. 
{\color{black}These anchors will receive nearby ML models from devices as they leave the network and transmit the latest global ML model to new devices as they join the network. In doing so, anchor devices improve ML training by enabling more devices to contribute to the training process in-between global model aggregations.} 
% We propose the concept of anchor devices in FL as edge devices that can receive the ML models of nearby devices and send the global ML model to new devices that enter the network. 
% In doing so, anchor devices can improve ML model training as they can gather the ML models of devices that leave the network and they enable devices that enter the network to begin ML training by sending them the latest global ML model. 
\subsubsection{$n$-hop cooperation} 
{\color{black}$n$-hop cooperation aims to broaden the scope of D2D cooperation, providing resource-scarce devices with greater access to resource-abundant ones through intermediary devices. Consequently, this technology can further enhance the resource and time savings introduced by single-hop D2D cooperation. This calls for novel optimization methodologies to characterize the benefits and trade-offs of $n$-hop cooperation.}
% Introducing and optimizing n-hop cooperation Furthermore, the cooperative aspect of PSL can be extended to the development of multi-hop data/model offloading, which can extend the distributed framework beyond those devices that can connectivity to the central server. 
\subsubsection{Heterogeneous privacy needs in D2D cooperation} 
Edge/fog devices may have heterogeneous privacy needs. For example, D2D connections may be allowed based on trust or familiarity. 
In such cases, devices often band together into cliques, which are private groups with a certain level of mutual trust.
{\color{black}Data transferring can be restricted to links between mutually trusted devices (e.g., smart devices such as a smartphone, laptop, and tablet of the same owner). 
In intra-clique cooperation, then, devices can share data without any restrictions, while, in inter-clique cooperation, devices may only be willing to share model parameters or insensitive data.
Furthermore, dataset stratification can be designed to separate data based on sensitivity, with restricted offloading of sensitive strata (e.g., personal health data or biometrics) among intra-clique devices and unrestricted offloading of insensitive strata (e.g., weather information).}

%%% uncomment interference management if we need another future work example
% \subsubsection{Interference management}
% While problems such as interference management have been widely studied in communications literature, the link between wireless interference on the success or timeliness of distributed machine learning at scale has yet to be properly investigated. 
% An integration of the joint problems of $\mathsf{PSL}$'s cooperation driven FL and the traditional communications problem of interference management can lead to novel formulations that determine optimal data and model routing in cooperative edge/fog networks. 

\subsubsection{Inclusion of devices with unlabeled data}
In practical edge/fog networks, some devices may have mostly or fully unlabeled datasets. 
% In federated settings, these devices are also likely to have non-i.i.d. underlying data distributions, so a global ML model trained by devices with labeled data may not be effective for devices with unlabeled data.
Standard FL neglects all these devices and obtains a global ML model by only engaging the devices with labeled datasets.
{\color{black}Through D2D cooperation, edge devices can share small quantities of data, labeled or unlabeled, to develop estimates of data distributions at devices with unlabeled datasets. 
This technique, termed unlabeled distribution estimation, will then involve determining unique combinations of ML models trained by devices with labeled datasets for use at devices with unlabeled datasets.}

\section{Network-aware D2S Cooperation} %Inclusion of Edge Servers} % Infrastructure} %}Towards D2S Collaboration} %Multi-Edge Server}
\label{sec:multi_server}
\noindent
Edge servers, especially in large-scale edge/fog networks, offer an untapped resource in standard FL. 
{\color{black}D2S cooperation aims to facilitate efficient utilization of these resources, e.g., by enabling devices with limited computation capabilities to transfer local training data to edge servers. In doing so, edge servers can leverage their powerful and efficient processors to improve energy efficiency and training delay of ML model training, thus enhancing the trade-off between energy consumption and ML performance.}
% D2S cooperation can unlock this potential by enabling devices to transfer their training data to edge servers, which can leverage their more powerful and efficient processors to provide energy efficient and faster ML model training, or use edge server to efficiently route ML models during aggregations. 
% Furthermore, D2S cooperation can use these edge servers to more efficiently route ML models during aggregations. 
In the following, we introduce a set of core technologies that enable effective D2S cooperation, and thereafter explain complementary future technologies to further enhance it. %on complementary technologies to further 

\begin{figure}[t]
\includegraphics[width=.48\textwidth]{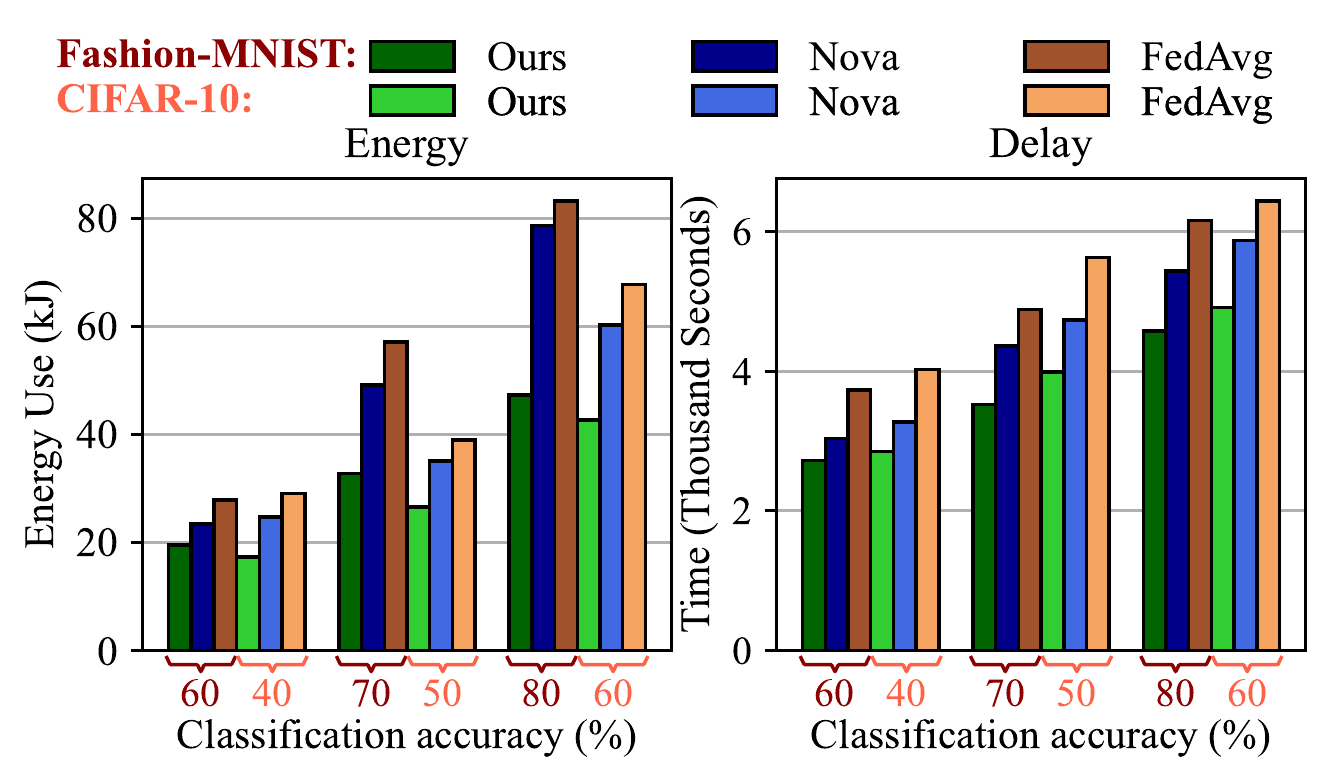}
\centering
\caption{{\color{black} On both Fashion-MNIST and CIFAR-10, our method based on  D2S collaborations enable both energy and time savings during the ML model training process.}}
\label{fig:D2S_result}
\vspace{-3mm}
\end{figure}

%Multi-Edge Server-Assisted Dynamic Federated Learning with an Optimized Floating Aggregation Point

\subsection{Core Techniques for Effective D2S Cooperation}
% \subsection{Multi-edge server-assisted dynamic federated learning}
Edge servers can have a diverse set of functionalities when assisting ML. They can act as computational resources to train ML models, as communication gateways to reroute ML models during global aggregations, and as the model aggregation points. 
% These use cases all involve server and edge device cooperation to improve the resource consumption of ML model training and the performance of the ML model, and we propose novel technologies to enable such cooperation. 
Starting at the computational level, we propose a novel technology called load balancing of data processing tasks, that relies on data offloading from edge devices to edge servers. Then, we propose efficient ML model parameter and data routing from edge devices to edge servers using base stations and network routers. Finally, we develop a concept called floating aggregation point, a method to optimize the selection of the aggregation server to save communication resources and minimize communication delay. 
% In standard FL, one edge server is selected as the central server by default, though this process has yet to be properly defined in literature. 
% Our recent methodology called cooperative edge-assisted dynamic federated learning ($\mathsf{CE-FL}$) explores the potential of using edge servers and network infrastructure to improve the base processes inherent to FL and extend FL to time-varying and non-stationary edge/fog networks~\cite{ganguly2022multi}. 
% In such edge/fog networks, not only are underlying data distribution at edge devices time varying but edge devices also have mobility, so that geographical proximity to edge servers and other network devices is time-varying as well.
% To these ends, $\mathsf{CE-FL}$ optimizes D2S collaborations throughout the network and dynamically re-adjusts the aggregation server at every aggregation, 
We present a visual summary of these new technologies in Fig.~\ref{fig:D2S}, and explain them in detail below.

\subsubsection{Load Balancing of Data Processing}
In standard FL, edge devices perform all of the computationally intensive ML model training tasks. We propose a novel technology of load balancing for data processing tasks in order to make use of the computational power at edge servers, similar to mobile edge computing frameworks in large-scale edge/fog networks~\cite{kaewpuang2013framework}. 
As part of load balancing, resource-constrained edge devices have the option to transfer a subset or all of their local data to nearby base stations (or road-side units), which through efficient data routing (another innovation which we explain next) relay the data to one or many edge servers. 
% Once these edge servers complete the ML model training, 

\subsubsection{Efficient Data and Parameter Routing}
To enable D2S cooperation, we propose efficient data and model parameter routing through the use of routers/switches as shown in Fig.~\ref{fig:D2S}. Through a combination of base stations and routers/switches, we can finely control the routing of data and ML models based on communication factors, such as channel congestion (a base station may be serving many users), and systems factors, such as computation power availability (an edge server may be running intensive data backups). 
This fine-grained control of data and ML model routing enables energy efficient and fast ML training. 
% networks to train ML models with greater time and energy efficiency. 

\subsubsection{Floating Aggregation Point}
In scenarios with many edge servers, we can improve resource efficiency by dynamically selecting the aggregation server. %the choice of aggregation server becomes an open question. 
Specifically, we propose floating aggregation point, a novel technology that adjusts the aggregation server based on changing edge/fog network properties. 
% To address this, we propose a novel technology called floating aggregation point, where the aggregation server changes over time based on the edge/fog network's properties.
As shown in Fig.~\ref{fig:D2S}, the choice of global aggregation server changes in response to devices' positions and dataset sizes, which influence the total communication/computation resource consumption for ML model training and parameter aggregation (uplink) and broadcasting (downlink). % as well as the communication delay at aggregation 
% The benefits of D2S collaboration and edge server inclusion become more apparent in time-varying and non-stationary networks, where flexible aggregation servers can lead to communication resource savings and reductions to network aggregation delay. 

% \subsubsection{D2S Cooperation Driven Optimization of ML Performance, Energy Consumption, and Delay}
% By combining the above three proposed technologies, we can jointly control D2S cooperation characteristics (e.g., data or ML model offloading ratios) to influence ML performance, energy consumption, and delay. 

{\color{black}We have taken initial steps towards formalizing such a D2S cooperation methodology in~\cite{ganguly2022multi}, including corresponding mathematical formulations and a proof-of-concept testbed implementation.
To demonstrate the potential benefits, we compare our method to FedAvg~\cite{kairouz2021advances} and Nova~\cite{wang2021novel} on two common benchmark datasets in FL literature: Fashion-MNIST (clothes) and CIFAR-10 (common objects) in Fig.~\ref{fig:D2S_result}. In this experiment, we consider a network of 20 edge devices and 10 edge servers training a two layer CNN. Other system parameters, such as the wireless channels and edge server links, can be found in~\cite{ganguly2022multi}.}
% {\color{black}To evaluate the above proposed techniques, we compare our method to FedAvg~\cite{kairouz2021advances} and Nova~\cite{wang2021novel} on two common benchmark datasets in FL literature: Fashion-MNIST (clothes) and CIFAR-10 (common objects) in Fig.~\ref{fig:D2S_result}. In this experiment, we simulate a network with 20 devices and 10 edge servers as it trains a two layer CNN. Other network parameters such as wireless channels and edge server links model are chosen similar to~\cite{ganguly2022multi}.} 
% {\color{black}We compare our D2S cooperation methodology against Nova for the same reasons as those for D2D cooperation.} 
This shows that CFL obtains substantial improvements over both baselines in terms of energy and time savings for the same target ML performance. 
% Together, our proposed technologies enable the control of D2S cooperation to jointly optimize ML performance, energy, and delay that is more effective than existing methods. 
% {\color{black} We have taken initial steps toward formalizing D2S cooperation methodology in~\cite{ganguly2022multi}, which includes mathematical formulation and an implementation methodology.}

% where we explain additional techniques to determine device-specific mini-batch sizes for ML model training, CPU clock frequency, and even heterogeneous training rounds per aggregation.

% We show an experimental example in Fig.~\ref{fig:D2S_result} of the combined effects of our D2S enabling technologies on two commonly used ML datasets - Fashion-MNIST (images of clothes) and CIFAR-10 (images of general objects) - relative to two baselines - FedAvg~\cite{kairouz2021advances} and FedNova~\cite{wang2020tackling}. 
% In Fig.~\ref{fig:D2S_result}, we see that D2S cooperation for FL can obtain substantial improvements both in energy and time savings for the same target ML performance. 

\subsection{Future Development of Complementary Technologies}
% While contributions of $\mathsf{CE-FL}$ are already substantial, 
D2S cooperation can be further extended  to enhance the existing trade-offs in FL and CFL depicted in Fig.~\ref{fig:main_diagram}. 
{\color{black}The following outlines a few open research directions:} 

% \subsubsection{Edge device mobility beyond the network}
% While the system model for $\mathsf{CE-FL}$ assumes that edge devices can move throughout the network, such devices may be able to move beyond the network~\cite{savazzi2021opportunities}. In such scenarios, the overall network-wide data distributions may shift dramatically. D2S collaboration must be further developed to account for such network boundaries.

\subsubsection{Non-stationary servers}
Given current trends leveraging unmanned aerial vehicles (UAVs) as mobile communication servers (e.g., at sporting events), a natural next step for D2S collaboration involves non-stationary edge/fog servers~\cite{savazzi2021opportunities}. 
{\color{black}Since mobile servers' locations can be controlled, efficient server placement methods can be pursued to improve data/model routing and offloading.
These methods should carefully investigate the trade-offs involved in server positioning. 
For example, placing a server near a dense neighborhood of devices may ease aggregation delay, but placing a server near a few resource-scarce devices may save more computation energy.}

\subsubsection{Joint D2D and D2S collaboration} 
Frameworks combining D2D and D2S can yield further benefits to ML performance, resource consumption, and time efficiency.
{\color{black}Such frameworks can enable simultaneous D2D and D2S data/model offloading. For example, as an edge device offloads data to another device, this secondary device could simultaneously can offload data to an edge server. This combined D2D and D2S cooperation ensures that (i) resource-abundant edge devices do not become overburdened, and (ii) resource consumption of data offloading (i.e., energy and delay) is optimized.}
% {\color{black} Such frameworks could enable simultaneous offloading, a new technique where both D2D and D2S data offloading occur at the same time.} 
% In simultaneous offloading, a resource-scarce edge device offloads data to a resource-abundant edge device which simultaneously offloads data to an edge server. This combined D2D and D2S cooperation would ensure that (i) resource-abundant edge devices does not become overburdened, and (ii) the data offloading delay is minimized. 

\subsubsection{Inclusion of unlabeled data} %Move contrastive learning here}
D2S cooperation can also be used to extend FL to edge/fog networks with fully unlabeled data, such as autonomous driving where camera-equipped cars take images without labels. One possible approach is to extended the well-known concept of contrastive learning~\cite{chen2020simple} to FL. %
Standard contrastive learning differentiates among datapoints in centralized settings via determining their similarities and differences. 
% determines similarities and differences among datapoints over a large centralized training dataset to differentiate between them. 
However, in federated settings, edge devices' datasets may simply be too small or lack sufficient data for standard contrastive learning to be effective. %tandard contrastive learning can have difficulty determining the key features of data on small and/or highly heterogeneous datasets, such as those at edge devices. 
% However, these existing methods apply poorly to federated settings as devices may have small and/or highly heterogeneous (i.e., non-i.i.d. distributed) datasets, which means there may be insufficient quantity and quality of local data to identify structural similarities and differences within the data. 
Through D2S collaboration, contrastive learning can be enabled in FL by leveraging edge servers as caches of data. %gathered across edge devices
As caches, edge servers can then supplement edge devices' local datasets with their cached data so that the compare and contrast steps of contrastive learning are feasible/effective in FL. 

\section{Conclusion} 
\label{sec:conc}
\noindent 
We proposed cooperative federated learning (CFL), a paradigm that extends the notion of cooperation in federated learning (FL) and unlocks the potential of edge/fog networks in the execution of distributed machine learning tasks. 
Through device-to-device (D2D) and device-to-server (D2S) cooperation, CFL counteracts the heterogeneity of edge/fog networks to improve ML model performance, energy efficiency, temporal sensitivity, and data/model privacy. We proposed novel technologies that enable efficient D2D and D2S cooperation in CFL. 
% edge devices can share data, ML models, and computational 
% When they involve device-to-device collaboration, devices can share data, model, and computational resources. We also show that the inclusion of device-to-server cooperation provides further opportunities for both energy efficient and faster ML model training. 
Finally, we illustrated how CFL can extend the frontiers of research in FL. % and, more generally, distributed ML. 

% \input{image_staging.tex}

% \newpage

\bibliographystyle{IEEEtran}
\bibliography{refs}

\vspace{-14mm}
\begin{IEEEbiographynophoto}{Su Wang (S'19)} is a Ph.D. candidate at Purdue University.
\end{IEEEbiographynophoto}
\vspace{-14mm}
\begin{IEEEbiographynophoto}{Seyyedali Hosseinalipour (M'20)} is an Assistant Professor at University at Buffalo (SUNY). 
\end{IEEEbiographynophoto}
\vspace{-14mm}
\begin{IEEEbiographynophoto}{Vaneet Aggarwal (SM'15)} is a Professor at Purdue University.
\end{IEEEbiographynophoto}
\vspace{-14mm}
\begin{IEEEbiographynophoto}{Christopher G. Brinton (SM'20)} is an Assistant Professor at Purdue University.
\end{IEEEbiographynophoto}
\vspace{-14mm}
\begin{IEEEbiographynophoto}{David Love (F'15)} is the Nick Trbovich Professor at Purdue University. 
\end{IEEEbiographynophoto}
\vspace{-14mm}
\begin{IEEEbiographynophoto}{Weifeng Su (F'18)} is a Professor at University at Buffalo (SUNY).  
\end{IEEEbiographynophoto}
\vspace{-14mm}
\begin{IEEEbiographynophoto}{Mung Chiang (F'12)} is the President of Purdue University. % and the Roscoe H. George Distinguished Professor at Purdue University. 
\end{IEEEbiographynophoto}

\end{document}